\documentclass[12pt]{article}

\usepackage{epsfig}
\usepackage{cite}
\usepackage{amsmath, amssymb, amsfonts}
\usepackage{xcolor}
\usepackage{latexsym}
\usepackage{graphicx}
\usepackage{soul}
\usepackage[colorlinks,bookmarks]{hyperref}
\hypersetup{pdfpagemode=UseNone, pdfstartview=FitH, linkcolor=blue,
            citecolor=red, urlcolor=blue}

\def\be{\begin{equation}}
\def\ee{\end{equation}}
\def\ba{\begin{eqnarray}}
\def\ea{\end{eqnarray}}

\def\bdm{\begin{displaymath}}
\def\edm{\end{displaymath}}
\def\la{~\mbox{\raisebox{-.6ex}{$\stackrel{<}{\sim}$}}~}
\def\ga{~\mbox{\raisebox{-.6ex}{$\stackrel{>}{\sim}$}}~}
\def\bq{\begin{quote}}
\def\eq{\end{quote}}

 at 10truept

\newcommand{\de}{\partial}

\renewcommand{\[}{\left[}
\renewcommand{\]}{\right]}
\renewcommand{\(}{\left(}
\renewcommand{\)}{\right)}
\newcommand{\vk}{\vec{k}}

\newcommand{\eps}{\epsilon}

\newcommand{\Om}{\Omega}

\newcommand{\Hz}{\textrm{Hz}}

\newcommand{\Mpc}{\textrm{Mpc}}

\newcommand{\Mpl}{M_{\mathrm{Pl}}}

\newcommand{\fnl}{f_{\mathrm{NL}}}

\newcommand{\bea}{\begin{eqnarray}}
\newcommand{\eea}{\end{eqnarray}}

\newcommand{\bi}{\begin{itemize}}
\newcommand{\ei}{\end{itemize}}

\newcommand{\beq}{\begin{equation}}
\newcommand{\eeq}{\end{equation}}
\newcommand{\beqa}{\begin{eqnarray}}
\newcommand{\eeqa}{\end{eqnarray}}
\newcommand{\mpl}{\Mpl}

\def\la{~\mbox{\raisebox{-.6ex}{$\stackrel{<}{\sim}$}}~}
\def\ga{~\mbox{\raisebox{-.6ex}{$\stackrel{>}{\sim}$}}~}

\newcommand{\calO}{\mathcal{O}}
\newcommand{\calR}{\mathcal{R}}
\newcommand{\vx}{\vec{x}}

\newcommand{\calL}{\mathcal{L}}

\def\ltap{\ \raise.3ex\hbox{$<$\kern-.75em\lower1ex\hbox{$\sim$}}\ }
\def\gtap{\ \raise.3ex\hbox{$>$\kern-.75em\lower1ex\hbox{$\sim$}}\ }
\def\gl{\ \raise.5ex\hbox{$>$}\kern-.8em\lower.5ex\hbox{$<$}\ }
\def\roughly#1{\raise.3ex\hbox{$#1$\kern-.75em\lower1ex\hbox{$\sim$}}}

\begin{document}

\thispagestyle{empty}
\begin{flushright}
December 2021\\
DESY 22-012 \\
\end{flushright}
\vspace*{.55cm}
\begin{center}
  
%{\Large \bf Perturbations and Signatures of Strongly Coupled}
%\vskip.3cm
%{\Large \bf Double Monodromy Inflation: a Redux}
{\Large \bf Very Hairy Inflation}

\vspace*{1cm} {\large Guido D'Amico$^{a, }$\footnote{\tt
damico.guido@gmail.com}, Nemanja Kaloper$^{b, }$\footnote{\tt
kaloper@physics.ucdavis.edu} and 
Alexander Westphal$^{c,}$\footnote{\tt alexander.westphal@desy.de}
}\\
\vspace{.2cm} {\em $^a$Dipartimento di SMFI dell' Università di Parma and INFN}\\
\vspace{.05cm}{\em Gruppo Collegato di Parma, Italy}\\
\vspace{.2cm}
{\em $^b$QMAP, Department of Physics and Astronomy, University of
California}\\
\vspace{.05cm}{\em Davis, CA 95616, USA}\\
\vspace{.2cm} $^c${\em Deutsches Elektronen-Synchrotron DESY, Notkestrasse 85,}\\
\vspace{.05cm}{\em D-22607 Hamburg, Germany}\\

\vspace{.5cm} ABSTRACT
\end{center}
We revisit the rollercoaster cosmology based on multiple stages of monodromy inflation. Working
within the framework of effective flux monodromy field theory, we include the full range of 
strong coupling corrections to the inflaton sector. We find that flattened potentials $V \sim \phi^p + \ldots$ 
with $p \la 1/2$, limited to $ N \la 25 - 40$ efolds in the first stage
of inflation, continue to fit the CMB. 
They yield $0.96 \la n_s \la 0.97$, and produce relic gravity waves
with $0.006 \la r \la 0.035$, in full agreement with the most recent bounds from BICEP/{\it Keck}. 
The nonlinear derivative corrections generated by strong dynamics in EFT  
also lead to equilateral non-Gaussianity $f_{NL}^{eq} \simeq {\cal O}(1) - {\cal O}(10)$, 
close to the current observational bounds. Finally, in multi-stage rollercoaster, an 
inflaton--hidden sector $U(1)$ coupling can produce a tachyonic chiral vector background, which converts rapidly 
into tensors during the short interruption by matter domination. 
The produced stochastic gravity waves are chiral, and so 
they may be clearly identifiable by gravity wave instruments like LISA, Big Bang Observer, 
Einstein Telescope, NANOgrav or SKA, depending 
on the precise model realization. We also point out that the current attempts to resolve the
$H_0$ tension using early dark energy generically raise $n_s$. This might  
significantly alter the impact of BICEP/{\it Keck} data on models of inflation.

\vfill \setcounter{page}{0} \setcounter{footnote}{0}

\vspace{1cm}
\newpage

\section{Intro}

In this article we complete the analysis of the observable predictions of interrupted
monodromy inflation models which we initiated in \cite{DAmico:2021vka}. Our goal is to obtain the
complete set of observables for monodromy models in strong coupling limit, which in addition 
to flattened inflaton potentials also include higher derivative
operators which renormalize the inflaton(s)' kinetic terms. While these corrections yield additional suppression
of the tensor to scalar ratio $r$, they tend to push the scalar spectral 
index $n_s$ closer to unity \cite{eva,DAmico:2021vka,Kallosh:2021mnu}.
That might tempt one to interpret the data as a serious obstacle for monodromy models, due to 
quite tight bounds on $r$ and $n_s$, $r \la 0.035$ and $0.96 < n_s \la 0.97$ \cite{bicep}. 
However, as we will show explicitly in this work,
such conclusions would be far too premature. On the contrary, when combined with the idea of
rollercoaster cosmology \cite{DAmico:2020euu}, monodromy models which include flattened potentials -- possibly
due to first-principle constructions \cite{McAllister:2014mpa,Dong:2010in} or generated by 
field theory strong coupling corrections \cite{Kaloper:2011jz,Dubovsky:2011tu,DAmico:2017cda}, or even
in the multifield models -- remain perfectly viable candidates to explain all the 
cosmological observations to date, while simultaneously generating 
new signatures that can be tested by the next generation of instruments. We should
also stress that the spectral index will be lower when many fields move in unison, simulating a single composite
inflaton \cite{Kaloper:1999gm,Kim:2006ys,Wenren:2014cga}. 
This is because different fields with nearly degenerate masses
fall out of slow roll at different times, inducing a small but potentially relevant correction to $n_s$. 

In the present work, the key input will be interrupting inflation 
at around $20-35$ efolds before the end, by the choice of the multifield potential. The interruption resets the
pivot at which the observables are generated, and normalized. In turn this pushes $n_s$ back to lower values and
restores consistency with the observations, while simultaneously keeping $r \la 0.035$
thanks to the potential flattening. Even if the first principles constructions generate steeper
potentials, the potential flattening induced by strongly coupled physics 
could be very efficient by itself, thanks to both the ``seizing" effect induced by wave-function
renormalization \cite{Dimopoulos:2003iy,Kaloper:2011jz,Dubovsky:2011tu,DAmico:2017cda}, and to 
the higher derivatives which further assist by reducing 
the speed of sound of the perturbations \cite{DAmico:2017cda}. 
However, in the latter case the nonlinear derivative operators simultaneously 
generate non-Gaussianities that could disrupt the
CMB if they are too large (such phenomena are automatically very suppressed at weak coupling). 
Suppressing non-Gaussianities yields a lower bound on $r$, about $r \ga 0.006$, for the
models with the flattened potentials $V \sim \phi^p + \ldots$, $0.1 \le p \la 1/2$, 
which are at the edge of the ``penumbra around the lamppost"
of the string theory motivated constructions at this time. 

Moreover, if the interrupted inflaton, which is
an axion-like field, also couples to a dark photon with the decay constant near the scale of inflation,
it will generate very large dark photon chiral fields. These in turn source chiral gravity waves, 
that are frozen out at superhorizon scales during the second stage of inflation, lasting
another $20-35$ efolds \cite{DAmico:2021vka}. These gravity 
waves present a very strong signal which could be detected
by future instruments like, for example, LISA, Big Bang Observer, 
Einstein Telescope, NANOgrav or SKA, depending on the precise details of the models
and the scales which they require, or rather, where the stages of inflation are interrupted.
This makes our specific subclass of strongly coupled 
EFT monodromy models particularly interesting, since 
their signatures are within reach of the multiple near-future observations. 

To proceed, in the next section we will discuss how to include the higher derivative corrections 
induced at strong coupling, and reanalyze their effects on the background dynamics of inflation
and on perturbations. Solving the equations numerically, we will fit the results against the most 
recent BICEP/{\it Keck} data \cite{bicep}, confirming the assertions above that they fit CMB, interpreted within 
the $\Lambda$CDM ``concordance model" of late cosmology, perfectly, with 
$0.96 \la n_s \la 0.97$ and $0.006 \la r \la 0.035$, when the first stage of inflation is 
interrupted a few decades of efolds before the end. 

We will then show how the lower bound on $r$ comes about due to the combination of non-Gaussianities 
induced by the higher derivatives terms and the assumption that $p \ge 0.1$. We impose this condition 
to stay near the range of parameters that can be related to explicit string theory motivated corrections. We will
also use a toy model of a theory with higher derivative operators to illustrate just 
how significant these effects can be. 

In Sec. 3, we will revisit the mechanism for chiral gravity wave production, described in \cite{DAmico:2021vka}.
We will explain that the results transfer straightforwardly from the previous case, 
where we neglected the higher derivatives, 
because by the time the first stage of inflation ends the inflaton sector is in weak coupling.
We will see that for powers $1/2 \ge p \ge 0.25$, the interruption pivot occurs at $\sim 20-35$ efolds 
of inflation before the end. Hence for those cases the produced primordial gravity waves are in the range of LISA
or DECIGO. If powers are lower, $0.25 \ge p \ge 0.1$, inflation may need to end sooner implying that 
the produced chiral gravity waves could have larger wavelength, and might be in the range of SKA 
or NANOgrav, or other instruments. 

Finally, we will also stress a rather curious and unexpected connection between the problem
of constraining the early universe cosmology and inflation, and the determination of the late universe value of
the Hubble parameter $H_0$. The $\Lambda$CDM ``concordance cosmology" may be facing a serious 
challenge from the intriguing discrepancy in determining $H_0$ using CMB from that one using supernovae 
\cite{Bernal:2016gxb,Verde:2019ivm,DiValentino:2018zjj,Poulin:2018cxd,Freedman:2019jwv,Niedermann:2019olb,
Niedermann:2020dwg,DiValentino:2021izs,Schoneberg:2021qvd}. 
The two values are $H_0 \simeq 68 \, {\rm km/sec/Mpc}$ and $H_0 \simeq 73.3 \, {\rm km/sec/Mpc}$, 
and according to the latest
analyses this leads to a $5 \sigma$ disagreement between the two \cite{Riess:2021jrx}. 
An interesting interpretation of this problem
is that plain vanilla $\Lambda$CDM may not be right, and a leading contender which might reconcile the two
values of $H_0$ might be the introduction of a new material in the universe commonly 
dubbed ``Early Dark Energy" (EDE) 
\cite{Poulin:2018cxd,Niedermann:2019olb,Niedermann:2020dwg}. Refitting the CMB to 
the late data however seems to require raising the primordial value of
$n_s$. This therefore directly influences the bounds on the early inflation, despite a naive expectations that the
two regimens, the early and the late universe, should be `decoupled'. This is not a 
deep conceptual issue, however, but
is a potentially serious practical aspect of early universe imprint 
contamination by late universe evolution, and is a familiar
issue from early forays into CMB fits \cite{lasenby,Kinney:2001js,Mukhanov:2003xr,Keeley:2020rmo}. 
We will not dwell on possible resolutions of the $H_0$ tension here beyond pointing 
out that the implications for the $r$ versus $n_s$ bounds may be quite dramatic, 
broadening the range of values for $n_s$, and ruling ``in" models
of inflation alleged to have been excluded, and severely constraining 
other models currently touted to be ``in good standing".

Either way, the jury is still out, and we are looking forward to whatever exciting news come our way.

\section{Flexing the BICEP of the Double-coaster}

As in \cite{DAmico:2021vka}, we will work here with a simple two-field model 
which supports a two-stage inflation that
is interrupted before it realizes $60$ efolds. We note that in more general cases, the early universe could involve
more than just two stages of inflation, with multiple inflatons which can leave signatures at many different 
scales \cite{DAmico:2020euu}. We will set the details of this more general and interesting possibility 
aside in this work 
for the sake of simplicity. However we will invoke them later, when outlining a range of predictions. 
It would be interesting to explore such scenarios in further detail. 
In the two field case, during the interruption, the first inflaton field oscillates, such that 
the effective equation of state of the 
universe in this epoch is $w\simeq 0$. This reproduces the dynamics considered 
in the past in 
\cite{Kofman:1985aw,Kofman:1986wm,Adams:1997de,Starobinsky:1985ibc,Polarski:1992dq,Peter:1994dx,
Cicoli:2014bja,Zhou:2020kkf,Tasinato:2020vdk,Braglia:2020eai}, 
albeit at a different pivot
relative to the beginning of inflation. Models 
like this are readily found in monodromy constructions 
\cite{Silverstein:2008sg,McAllister:2008hb,Kaloper:2008fb,Flauger:2009ab,Kaloper:2011jz,McAllister:2014mpa}, 
which typically involve more than one field, 
where the fields are separated by with small mass hierarchies. 

We recall that there is a range of possible effects which can flatten the potentials without and with strong coupling. 
For example, at weak coupling \cite{McAllister:2014mpa,Dong:2010in} integrating 
out fields heavier than the inflaton may lead to potential 
flattening as in \cite{Dimopoulos:2003iy}, and the theory may be under 
control even in weak coupling. In strong coupling, such effects may be 
ubiquitous in flux monodromy models \cite{DAmico:2017cda,DAmico:2021vka}, where
the fact that the inflaton is a magnetic dual of a longitudinal helicity of a 
massive $3$-form potential gauge theory \cite{Kaloper:2008fb,Kaloper:2011jz,Kaloper:2016fbr}
allows gauge symmetry to ``cherry pick" 
the right kinds of higher dimension operator corrections, which, while perserving gauge symmetry,
generically flatten the potential. In fact, just allowing several fields
simultaneously in slow roll, as in assisted inflation/N-flation 
\cite{Liddle:1998jc,Kanti:1999vt,Dimopoulos:2005ac}, could push
the spectral index down \cite{Kaloper:1999gm,Kim:2006ys,Wenren:2014cga}.

Bearing all of the previous elaboration in mind, 
we will take an `ignoble' approach and develop a model based only on
EFT+gauge symmetries. This means, we will take each inflaton to 
be an axion-like field, which is a magnetic dual of a
massive $3$-form potential gauge theory. For the most part we will forego 
the details of the UV completion which we do not have,
and using gauge symmetries and naturalness push forward to show that the EFT, when allowed to employ strong
coupling effects (just) below its cutoff, may yield a successful model of 
inflation \cite{Kaloper:2016fbr,DAmico:2017cda}. 
The low energy theory, in effect, is really a theory of gravitating superconductor, advocated in \cite{Kaloper:2016fbr}. 
With this in mind, we proceed with retaining the potential we used in \cite{DAmico:2021vka},
\be 
    V(\phi_1, \phi_2) = M_1^4 \[ \(1 + \frac{\phi_1^2}{\mu_1^2}\)^\frac{p_1}{2} - 1 \]
    + M_2^4 \[ \(1 + \frac{\phi_2^2}{\mu_2^2}\)^\frac{p_2}{2} - 1 \] \, .
\label{eq:Vdouble}
\ee
As before, the scales $\mu_1, \mu_2$ normalizing the fields are both $\calO(0.1 \Mpl)$. 
We take $M_2 \la M_1 \simeq M/\sqrt{4\pi}$, where $M/\sqrt{4\pi}$ is the strong coupling 
scale of the theory and $M$ the UV cutoff. We imagine that it is set by the mass of the lightest heavy field 
which was integrated out to obtain the effective potential (\ref{eq:Vdouble}). 

Both $\phi_i$ are axions arising from truncating $p$-form gauge potentials in 
string theory constructions, either as directly $p$-form components in compact directions or as
the magnetic duals of the residual $4$-forms\footnote{The distinction is important in understanding the 
origin of higher derivative operators later on.}. This immediately connects masses and axion decay constants, 
$\mu_i\sim f_i$. The latter are further bounded by $f \sim M_{Pl}/(M_s L)^q\lesssim M_{Pl}$ 
where $L$ is the size of the compactification cycle giving rise to the relevant axion, and $M_s$ the 
string scale \cite{Banks:2003sx,Svrcek:2006yi}. As a result, $f_i$ is typically 
of the order of $10^{-2}M_{Pl}\lesssim f\lesssim M_{Pl}$, justifying our choice of $\mu_i$. 
The scales $M_i$ typically arise 
from warping effects or dilution of energy densities with inverse powers of extra dimension volumes 
(see e.g. section 4.1 in~\cite{Dias:2018koa} for a summary), which are either power-law or exponentially 
sensitive to the microscopic parameters in a string compactification. Finally, the axions $\phi_i$ arise 
from two mutually sequestered sectors of a given model, only interacting via gravity, 
and so generically $M_1\neq M_2$. This also explains why we ignore the mixing between 
the two axions. Without loss of generality we take $M_2 \la M_1$.

Next, in this work we will also include the higher derivative operators which correct the kinetic terms,
which as we stressed are unavoidable in the strong coupling regime by naturalness of the perturbation
theory \cite{DAmico:2017cda}. In this regime of the theory, the only way to avoid some 
of these operators, short of fine tuning, is if there is some symmetry prohibiting them. 
Note that in weak coupling, below the cutoff, in explicit constructions
such operators need not play a significant role, while the potential may nevertheless be flattened 
\cite{McAllister:2014mpa,Dong:2010in}, due to the features of the UV completion of the direct string construction. 
Our point here is that even without knowing the full details of the string construction and 
the precise realization of the
UV completion, as long as the low energy EFT has gauge symmetries as in flux monodromy \cite{Kaloper:2016fbr},
some flattening of the potential will happen in strong coupling when the gauge invariant higher dimension
operators are included. As we already stated, this may well be the most conservative approach 
to the realization of models
with flattened potentials, and more efficient means may be found. However, in our view the merit of our
approach is that it {\it accommodates} the inflationary evolution -- just like the London theory of superconductivity
accommodated superconducting phenomena decades before BCS. 
The higher dimension operators of interest take the form \cite{DAmico:2017cda}
\be \label{correctionsscalarnorm}
{\cal L} \ni \sum_{k\ge 1, \, l \ge 1}  c_{k,l}\frac{(m_j\phi_j)^{l} }{{2^k k! l! (\frac{M^2}{4\pi})^{2k+l-2}}}
(\partial_\mu \phi_j)^{2k}  \, ,
	\ee
where $m_i^2 \simeq p_i^2 {M_i^4}/{\mu_i^2} < M^2$ are the effective inflaton masses in the flattened regime. 

We take the powers $p_i$ to be small, $0.1 \la p_i \le 1$, reflecting the flattening of the 
potentials due to the interactions 
with heavy fields which are integrated out \cite{Dong:2010in,DAmico:2017cda,DAmico:2021vka}. 
In strong coupling regime, we can see how such terms arise by imagining higher dimension
operators $\sim (\frac{4\pi \phi_i}{M})^q (\partial \phi_i)^2$ renormalizing the kinetic 
terms in the strong coupling regime 
$1> {\phi_i}/{M} \ge 1/4\pi$, and canonically normalizing the fields.  
Such powers can be realized in string constructions
`under the lamppost' which therefore may be under 
control\footnote{To have a valid EFT we must have a good description
of its vacuum. As field values change, and couplings run, the perturbative 
vacuum itself will evolve. In UV complete frameworks
one may have a full view of this evolution. In perturbative EFT we do not always have the benefit of such insight.}.

The effective action for 
monodromy  composed of (\ref{eq:Vdouble}) and (\ref{correctionsscalarnorm}) can 
be symbolically `resummed\footnote{We set aside the convergence issues, and  
take the `sum' to be an asymptotic series.},'  yielding for each inflaton  
\be \label{kinflation}
{\cal L}_i =  K\Bigl(\phi_i, X_i \Bigr) - V_{eff}(\phi_i)  =  
\frac{M^4}{16\pi^2} {\cal K}\Bigl(\frac{4\pi m_i \phi_i}{M^2}, \frac{16\pi^2  X_i}{M^4} \Bigr) 
- \frac{M^4}{16\pi^2} {\cal V}_{eff}(\frac{4\pi m_i \phi_i}{M^2})  \, ,
\ee
where $X_i \equiv - (\partial_\mu \phi_i)^2$, the 
functions ${\cal K}, {\cal V}_{eff} = 16\pi^2 V_{eff}/M^4$ have Taylor coefficients
$\sim {\cal O}(1)$, and 
we normalized the expansion using the strong coupling scale $M$ (which can be thought of as the 
mass of the lightest particle coupled to $\phi_i$ which was integrated out. 
The ${\cal K}$ is the kinetic function of the theory,
which in the weak coupling reduces to $(-1+ \ldots) (\partial \phi)^2/2$.

\begin{figure}[ht]
    \centering
    \includegraphics[scale=.85]{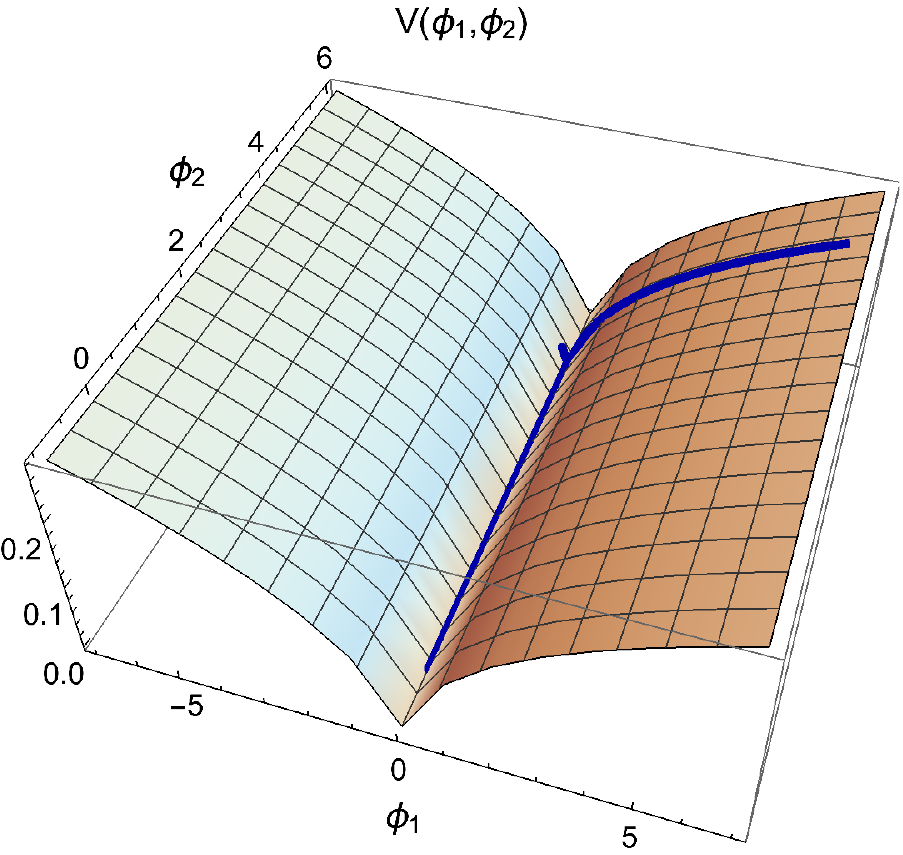}
    \caption{Two-field potential $V(\phi_1, \phi_2)$ for the model in eq.~\eqref{eq:Vdouble}.
    Here $M_1 = 0.5 \Mpl$, $M_2/M_1 = 0.1$, $p_1 = 2/5$, $p_2 = 1$, $\mu_1 = \mu_2 = 0.5$. 
    The blue curve depicts a
    typical two-stage inflationary trajectory, where the field $\phi_1$ 
    slides down the slope first, oscillates while decaying 
    near the bottom of the ``gutter", and then $\phi_2$ starts to move along the ``gutter". }
    \label{fig1}
\end{figure}

We have chosen to decouple 
the two axion sectors from each other for simplicity, and as a result the early inflationary trajectory 
and the late inflationary dynamics become two separate stages. Thus in this limit 
we can study the dynamics of inflation 
as a sequence of two consecutive single-field stages. As noted above, this could 
happen as a leading approximation in the
case when the two sectors mix only gravitationally. As in \cite{DAmico:2021vka} we view this as 
an interrupted N-flation \cite{Dimopoulos:2005ac},
with a larger mass gap between the two inflatons, where one 
dominates early on, and drops out of slow roll well before the other one. 
Each effective action ${\cal L}_i$ is precisely the action of the k-inflation model of 
\cite{ArmendarizPicon:1999rj,Garriga:1999vw}, with a single dimensional normalizing scale,
and the dimensionless coefficients of ${\cal O}(1)$ up to combinatorial factors set by normalizing 
Feynman diagrams. At strong coupling, where the theory (\ref{kinflation}) 
is driven by observations, it is this action which should be deployed to compute 
inflationary background and observables \cite{DAmico:2017cda}. In Fig. \ref{fig1}, we 
illustrate the inflationary trajectory on the potential (\ref{eq:Vdouble}). In the present case, the 
segments of the trajectory are also affected by the higher derivative 
operators (\ref{correctionsscalarnorm}) which we retained.

We also note that an even more extreme reduction of $r$ can occur if
the potential ${\cal V}_{eff}$ at large $\phi$ plateaus. In a sense, the change of 
the potential from convex to concave as strong coupling
regime sets in may be viewed as a beginning of such plateauing. In some cases where additional control tools are
present, namely AdS/CFT, and the corrections are enhanced by the presence of many flavors, 
such potentials could be constructed 
\cite{Dubovsky:2011tu,nomura}. In the example \cite{Dubovsky:2011tu}, the resulting effective potential is 
${\cal V}_{eff} \sim (1- \frac{1}{1+ (4\pi c m \varphi/M^2)^p})$. In this case 
the tensor to scalar ratio is very small, $r \sim {\rm few} \times 
10^{-4}$ \cite{Dubovsky:2011tu,DAmico:2017cda}. We will, 
however, not explore the detailed dynamics of such models here. 

The next step is to recalculate the spectrum of perturbations on large scales. 
As before we solve the equations of motion for the inflaton and calculate the 
scalar spectral index $n_s$ and the tensor-to-scalar ratio 
$r$ at $N_e$ efolds before the end of the first stage, but now we include the 
effect of the higher derivative operators. This forces us to resort to numerical integration, 
and we will only present the result for the first stage 
of inflation since this is when the fluctuations distorting the CMB are generated. At this stage 
we can approximate the dominant contribution to the effective potential during the first stage by
\be 
    V_{eff}(\phi_1) = 
    M_1^4 \[ \(1 + \frac{\phi_1^2}{\mu_1^2}\)^\frac{p_1}{2} - 1 \] \simeq M_1^4 \[ \(\frac{\phi_1}{\mu_1}\)^{p_1} - 1 \] \, .
\label{eq:Vdoublesin}
\ee
which in the plot of Fig. \ref{fig1} corresponds to fixing $\phi_2$ at 
some value and rolling down the hill towards the ridge 
at a $\phi_2 \simeq {\rm const.}$, thanks to $\phi_2 \gg \mu_2$. Here, clearly, $M_1 \simeq M/\sqrt{4\pi}$. 

In the absence of higher derivative operators
in \cite{DAmico:2021vka} we could then use the slow roll parameters 
$\eps_V = \partial_\phi V_{eff}(\phi_1)^2/(2 V_{eff}(\phi_1)^2)$ 
to both control the numerical integration of the dynamics based on (\ref{eq:Vdoublesin}) in slow roll regime, and to
compute the observables $n_s - 1 = 2 \eta_V - 6 \eps_V$, $r = 16 \eps_V$ at 
a value of $\phi_1$ when the first stage 
of inflation ends, which gives 
\be
N_e = \int_{\phi_1}^{\phi_1(end)} d\phi \, \frac{H}{\dot \phi} 
\simeq \(\frac{\phi_1^2}{2p_1 M_{Pl}^2}\)\[1 - \frac{2}{2-p_1} \(\frac{\mu_1}{\phi_1}\)^{p_1}\] \, 
\label{efolds}
\ee
efolds before the end of the first inflationary stage. That generated the primordial tensor modes with 
$r$ in the range $0.02 \la r \la 0.06$ \cite{DAmico:2021vka}. 

As noted \cite{Kaloper:2014zba,DAmico:2017cda}, 
higher derivative operators help reduce $r$ further. In this regime it is convenient to 
compute the cosmological observables using the k-inflation 
approach of \cite{Garriga:1999vw,Gruzinov:2004jx,Chen:2006nt}.
Importantly, the speed of sound of the scalar perturbations $c_s$ is different from unity, 
so that  the tensor-scalar ratio is $r = 16 c_s \eps$, and  $\eps$ is the usual slow-roll parameter. In this case 
the inflationary consistency relation is $r = - 8 c_s n_T$, where $n_T$ is the tensor spectral index.
What happens is that the higher derivatives improve the slow roll 
in the scalar field equation, without affecting the gravitational field equations.
As a result, the noted modification of the consistency condition follows.
Furthermore, the nonlinear terms in the fluctuations \cite{Gruzinov:2004jx,Chen:2006nt} 
may yield non-negligible equilateral non-Gaussianities.
We will also compute those numerically to show how the non-Gaussianities  limit $r$ from below. 

To do this, using the calculational framework of k-inflation, we 
restrict to the subcase when inflation is slow roll potential-driven, and
require that higher derivative terms dominate over the quadratic ones both 
in controlling the background and the perturbations. 
For simplicity, we will present below only the equations for the truncation of 
the kinetic energy function to only quartic terms, 
$K = {\cal Z} X_1 + \tilde {\cal Z} \frac{16\pi^2 X_1^2}{M^4}$. 
We will also consider the sextic derivative terms as an additional
example, motivated by direct string theory constructions. Both of these 
`trans-quartic' cases suppress $r$ more than the quartic truncations.
But in weak coupling they can go only so far. 

\section{An Interlude with High Harmonics, for Strings}

Before we proceed with computing the observables for the nonlinear 
theory we have set up so far, let us pause and consider in more
detail the origin of the nonlinearities, and in particular the higher derivative terms. 
The higher derivative operators may arise in
the weak coupling limit, in which case their coefficients may be calculable in perturbation theory. 
Generically this means the derivative-dependent
part of the effective action will be truncated to a few leading terms 
(as noted in the perturbative analysis in \cite{Kaloper:2014zba}, for example). 
The leading order terms are $\propto (\partial)^4$. However, although 
such terms can indeed arise already at weak coupling in string theory,
their impact is limited. In this section we stipulate what one may expect  in 
the controlled regime `under the lamppost', below the 
strong coupling scale. Our aim is to outline a `perturbative' boundary in the 
observable phase space, and to show by example how
strong coupling effects can enhance the observable signals. 

Which of the two `trans-quartic' cases mentioned above depends on the 
nature of the underlying monodromy inflaton candidate. If the inflaton scalar 
arises as a position modulus of a mobile D-brane acquiring 
monodromy e.g. via moving along a non-trivial fibration of the compactification space, 
its kinetic term may pick up higher-
derivative corrections to all orders from the DBI action of the moving D-brane 
generating the inflaton scalar potential as well. As a guiding 
example, we look e.g. at the D7-brane monodromy inflation model in \cite{Hebecker:2014eua} 
where the D7-brane position in the transverse 
extra dimensions is the inflaton candidate. F-theory there allows the authors 
to describe the D7-brane position as part of the 
elliptic Calabi-Yau 4-fold complex structure moduli. Therefore, the 
7-brane position acquires its two-derivative kinetic term from a Kahler potential 
in the effective 4D supergravity.

However, at least in the perturbative type IIB string theory limit of F-theory, this 
two-derivative kinetic for the D7-brane position is ultimately obtained via matching 
the expansion of the full DBI action of the D7-brane effective action in~\cite{Jockers:2004yj} 
up to two-derivative order. Hence, in the type IIB limit we expect the full kinetic term of the 
D7-brane position scalar inflaton candidate for the model of~\cite{Hebecker:2014eua}  
to take the complete DBI form. We leave a full discussion of these arguments for future work, 
but note that they provide plausibility for the appearance of the full infinite higher-derivative 
kinetic term series of DBI inflation in brane monodromy inflation setups of schematic type 
$\sqrt{1-\alpha' \dot x_\bot^2}$ such as seen e.g. in~\cite{Silverstein:2008sg}, as they rely 
just on the structure of the DBI action as the universal part of the perturbative effective description of D-branes.

In contrast, if inflation arises as axion monodromy from a bulk $p$-form string axion, this 
will not acquire contributions to its kinetic term from branes and thus no DBI-type kinetic term. 
Instead, the supersymmetric completion of the leading higher-order $\alpha'^3R^4$~\cite{Becker:2002nn} 
curvature correction is expected to generate corrections to the kinetic term of bulk closed string 
axions up to sextic order from terms containing at least one power of curvature. This expectation 
arises from considering the SUSY completion of the type II $\alpha'^3R^4$ correction. While this 
completion in its full form is unknown as of today, the bosonic sector of the completion, as  
sketched in e.g.~\cite{Conlon:2005ki}, is expected to contain terms reading schematically as
\begin{equation}
\Delta{\cal L}_{\alpha'^3R^4}^{SUSY,bos.}
\supset {R^3F_p^2, R^3|F_p|^2, R^2F_p^4, R^2|F_p|^4, RF_p^6, R|F_p|^6} \quad.
\end{equation}
The non-vanishing invariant tensor contractions among them clearly generate axion kinetic 
term corrections up to sextic order arising from the powers of $|F_p|^2$. Moreover, the possible 
invariants generating these higher-order axion kinetic terms may involve topological 
invariants of the compactification space arising from integrating over powers of curvature on 
the internal space. These topological invariants, like e.g. the Euler characteristic, may 
lead to sizable numerical enhancements of the higher-order 
axion kinetic terms, providing a rationale for their mass 
scales to scan a range of values relative to the Planck scale.

The higher-derivative axion kinetic term series inferred from the 
SUSY completion of the $\alpha'^3R^4$ terminates at a finite order of derivatives 
at this order in $\alpha'$. String theory is expected to generate corrections 
beyond ${\cal O}(\alpha'^3)$ in an infinite series. This is of course a string theory avatar of our
strong coupling EFT argument. Hence, at higher orders in  $\alpha'$ 
there will potentially exist invariants involving powers of the $p$-form field strengths 
producing higher-derivative axion kinetic terms beyond the maximal order generated 
at ${\cal O}(\alpha'^3)$.\footnote{We thank Timo Weigand for discussions concerning this point.}
But in weak coupling these terms will be small. The closed string sector 
higher-order $\alpha'$-corrections appear after reduction to 
4D suppressed by powers of the inverse compactification volume~\cite{Conlon:2005ki}. 

Hence, barring an appearance of numerically very large topological numbers governing these 
higher-order $\alpha'$-corrections, the higher-derivative axion kinetic terms potentially produced 
by them will have volume suppressed small coefficients relative to the higher-derivative axion 
kinetic terms appearing at ${\cal O}(\alpha'^3)$. In this sense the series of higher-derivative 
axion kinetic terms with potentially sizable coefficients appearing via dimensional reduction 
from the 10D closed string  sector will effectively terminate at sextic order. In other words, 
to get larger effects one must approach the limits of the validity of the compactified theory,
as expected from the generic EFT consideration. 

Again, we stress that the above observations about (ir)relevance of  
higher derivative terms are features of the weak coupling limit of the
theory, well below the relevant cutoff of the inflationary EFT, and under the assumption that 
the Hilbert space of states is largely unaffected by the background evolution of fields and couplings. 
As we approach the limits of the inflationary EFT, by cranking up the field values and
couplings, the irrelevant operators, including the higher derivative ones, will become more
influential. To account for those effects, we must be careful when comparing 
the higher-derivative axion kinetic terms 
arising from dimensional reduction of the 10D higher-derivative $p$-form field strength 
powers to the higher-derivative axion kinetic terms outlined in~\cite{DAmico:2017cda}. 
It is here where the origin of the low energy axions is important. 
The point is that simple `derivative accounting' may be misleading, which can be seen
as follows. Consider a low energy EFT axion, which acquires 
its quadratic potential via mixing with the 4-form field strength in 
the 4D effective flux monodromy description 
of~\cite{Kaloper:2008fb}, and is a pseudo-scalar magnetic dual of a perturbative 
$p$-form string axion, that comes from dimensional reduction. At two-derivative level 
all the known string axion and brane monodromy inflation mechanisms 
can be dualized into this 4D effective 4-form description \cite{Marchesano:2014mla}. 
This procedure is useful since it makes the hidden gauge symmetries of the theory manifest,
providing the tools to control the strong coupling regime. However, it also shows that once the dimensionally 
reduced $p$-form string axions acquire higher-derivative kinetic terms, it becomes quite non-trivial to 
match these to the description of the higher-derivative kinetic terms of the dual axion of 
the 4D effective 4-form theory, as encoded in couplings of the type 
$(A_3-d B_2)\wedge \star F_4$ with $B_2$ being the 4D dual 2-form gauge potential 
describing the axion degree of freedom from the compactified string model. In a nutshell, 
dualization is a canonical transformation \cite{Dvali:2005an,Kaloper:2016fbr}, and so in perturbation theory, 
the derivatives of the axion are mapped to the powers of the 3-form mass term and vice versa. This is
simply the consequence of the fact that a canonical transformation exchanges generalized coordinates
and momenta, $(q,p) \leftrightarrow (-P,Q)$, while preserving the commutation relations 
and the Hamiltonian \cite{Dvali:2005an,Kaloper:2016fbr}. The full ``resummation" of the 
perturbative duality maps also ``dresses" up the Hilbert space vacuum, very similarly to what 
happens with the BCS vacuum below the critical temperature, when the condensate forms. 
Thus it is perfectly plausible that a theory with higher derivatives on 
one side may appear as a totally non-derivative
theory on the other side. 

A flavor of the non-trivial duality matching is already visible in the 4-form description of 
hybrid axion monodromy in \cite{Kaloper:2020jso}, where the nontrivial issues concerning the
selection of the vacuum are noted. Here, we set such a full duality matching at higher-derivative order 
aside given the uncertainties with the perturbative dimensionally reduced 
$p$-form string axion sector at higher derivatives. 
We do expect that upon performing the duality match the finite order of the higher-derivative axion 
kinetic terms expected for a closed string axion at ${\cal O}(\alpha'^3)$ will 
translate to a matching finite higher-derivative order of the kinetic terms for 
the dual 4D effective axion of the 4-form EFT, which suffices for our purposes. We should also note that
even when the strong coupling effects induce large higher derivative operators, there are still regimes
where initial conditions for classical evolution allow flattened potentials to dominate, thus `deactivating' the
higher derivative terms \cite{DAmico:2017cda}. This regime is also illustrated by our truncation of the
derivative expansion. Obviously, this means there could be other branches of solutions which 
we ignore here. 

\section{Variations for BICEP3}

Having set up the stage for the EFT of double monodromy, with or without strong 
coupling effects included, and highlighted its 
relationship to the UV completions in models where inflatons are axions or brane positions, 
we can now leverage the universality of the 4D EFT to study the axion inflation dynamics at 
strong coupling. The homogeneous field configurations satisfy $16\pi^2 X_1/M^4 \gtrsim 1$ 
and $16 \pi^2 V_{eff}/M^4 \gtrsim 1$. Truncating to the quartic time derivative terms for 
simplicity, as outlined above, the slow roll field equations 
are~\footnote{When solving numerically, we work with the full $K(X)$ 
function in eq.~\eqref{background} in slow roll approximation.}
\be
3 \mpl^2 H^2 = V_{eff} \, , ~~~~~~~~~~~~~  6H \dot \phi_1 \tilde {\cal Z} X_1 = - 4\pi M^2 m V_{eff}' \, .
\label{background}
\ee
There are evolutionary regimes for the fields depending on the relevance of the 
derivative terms relative to the potential.  A discussion
of options was given in \cite{DAmico:2017cda}. A case of interest to us here is where 
the higher derivative operators are both large in the EFT, and are excited, contributing to the background 
during slow-roll inflation\footnote{As we explained above, it is possible that large 
higher derivative operators are present in the effective action, but that the initial conditions at the observable 
stage of inflation are such that these are subleading to the quadratic derivatives,
and remain so in slow roll. We ignore this here.}. In this case the tensor power is suppressed thanks to both the 
flattened potential and the subluminal speed of sound of the perturbations, induced by the higher derivatives.
An important additional consideration is that the higher derivative terms induce 
nonlinearities which yield non-Gaussianities. The
bounds on non-Gaussianities, $f_{NL} \la {\cal O}(10)$ imply a {\it lower} bound on $r$.

To compute the perturbations on the background controlled by the slow roll equations (\ref{background}), we deploy 
the formalism of the inflaton EFT \cite{Cheung:2007st}, ignoring the mixing with gravity. 
This approximation is good at energy scales 
$E \gg \sqrt{\eps} H$ and $c_s^2 \gg \eps$, which apply in our case. 
The spectrum of perturbations is almost scale-invariant and has 
small non-Gaussianities, as observations require. In the gauge where the spatial metric is unperturbed, the field 
perturbation is $\phi_1(t, \vx) = \varphi_{0}(t + \pi(t, \vx))$ whereas the spatial
curvature perturbation is ${\cal R} = - H \pi$. Expanding Eq.~\eqref{kinflation} 
up to third order, and keeping only the terms lowest in derivatives,  
\be \label{Lpi}
S = - \int dt d^3\vec x \, a^3 \Mpl^2 \dot{H} \[ \frac{1}{c_s^2} \dot{\pi}^2 - \frac{(\de_i \pi)^2}{a^2}
+ \(\frac{1}{c_s^2} -1 \) \( \dot{\pi}^3 + \frac{2}{3} c_3 \dot{\pi}^3
- \dot{\pi} \frac{(\de_i \pi)^2}{a^2} \)
\]  \, .
\ee
Dropping the subscript $"1"$ from here on, and using our monodromy EFT \eqref{kinflation}, the speed 
of sound is $c_s^2 = \partial_X {\cal K} / (\partial_X  {\cal K} + 2 X \partial_X^2  {\cal K})$, and 
$c_3(1/c_s^{2}-1) = 2 X^2 \partial_X^3  {\cal K}/ \partial_X  {\cal K}$. 
So when $X > \frac{M^4}{16\pi^2}$, we find $c_s^2 \sim \frac{M^4}{32\pi^2 X} \frac{{\cal K}''}{{\cal K}'}$ 
and $c_3(1/c_s^{2}-1) = \frac{512 \pi^4 X^2}{M^8} \frac{{\cal K}'''}{{\cal K}'}$, 
or $c_3 = \frac{16\pi^2 X}{M^4} \frac{{\cal K}'''}{{\cal K}''}$.
The exact details of the theory when $y = 16\pi^2 X/M^4 > 1$, which set the magnitude 
of ${\cal K}$\ and its derivatives, depend on the UV theory governing the large-$y$ asymptotia. 

\begin{figure}[t!]
    \centering
    \includegraphics[scale=0.85]{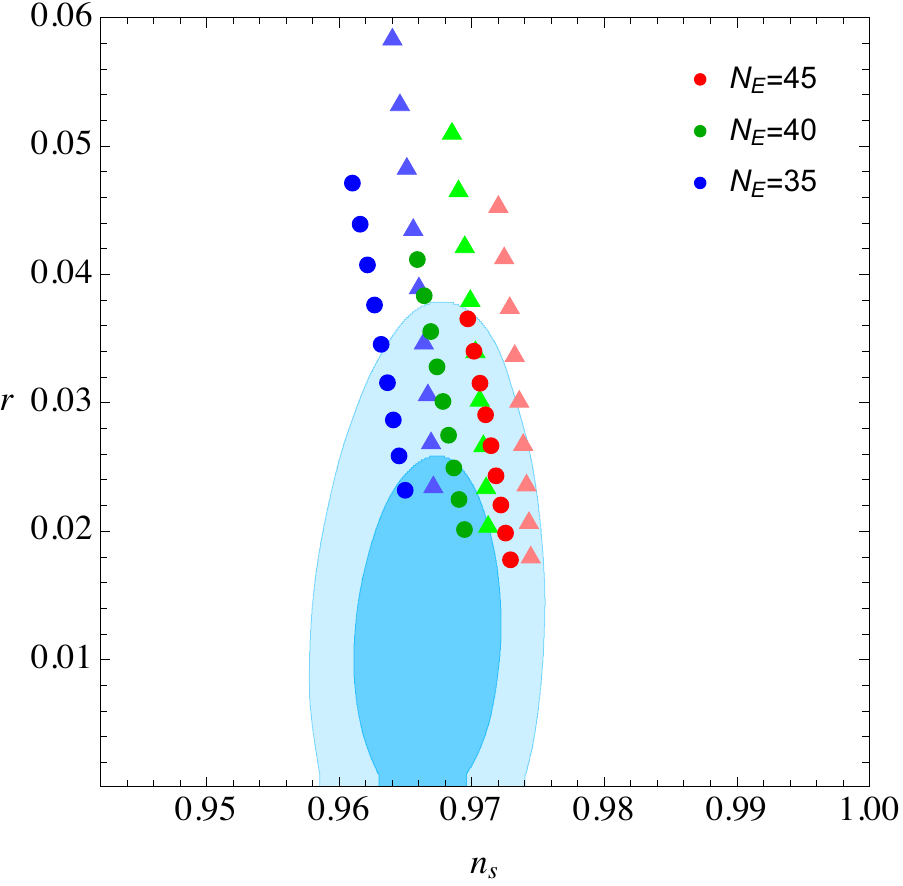}
    \includegraphics[scale=0.85]{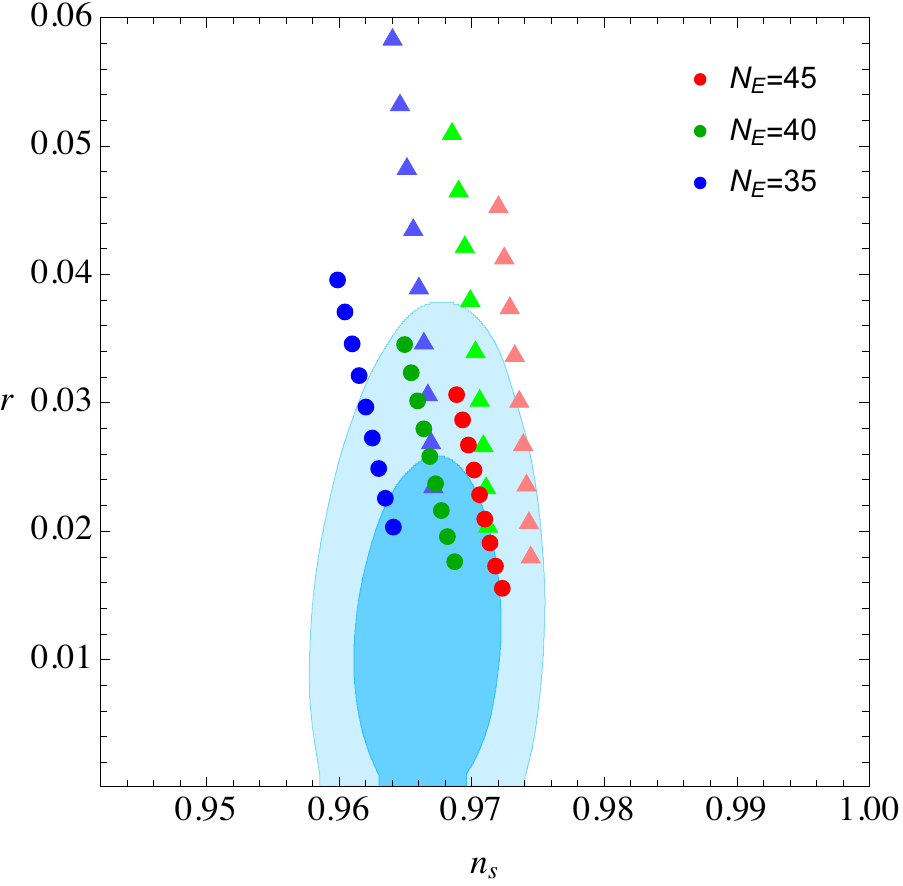}
    \caption{Tensor-to-scalar ratio versus spectral index for double monodromy 
    with quartic (\emph{left}) and sextic (\emph{right}) kinetic terms, compared to the data of~\cite{bicep}.
    In the left panel, $K(X) = X + X^2/M_4^4$, with $M_4 = 0.05$, while in 
    the right panel, $K(X) = X + X^3/M_6^8$, fixing $M_6 = 0.05$.
    We take $\mu_1 = 0.01$, and vary $0.1 \leq p_1 \leq 0.5$. The triangles 
    depict the quadratic kinetic term case $(c_s = 1)$, while dots are the predictions for the higher derivatives.
    A finite $M_4$ or $M_6$ will give a redder spectrum~\cite{Pedro:2019klo}, and a lower $r$.}
    \label{fig:nsrplot}
\end{figure}

\begin{figure}[t!]
	\centering
	\includegraphics[scale=0.85]{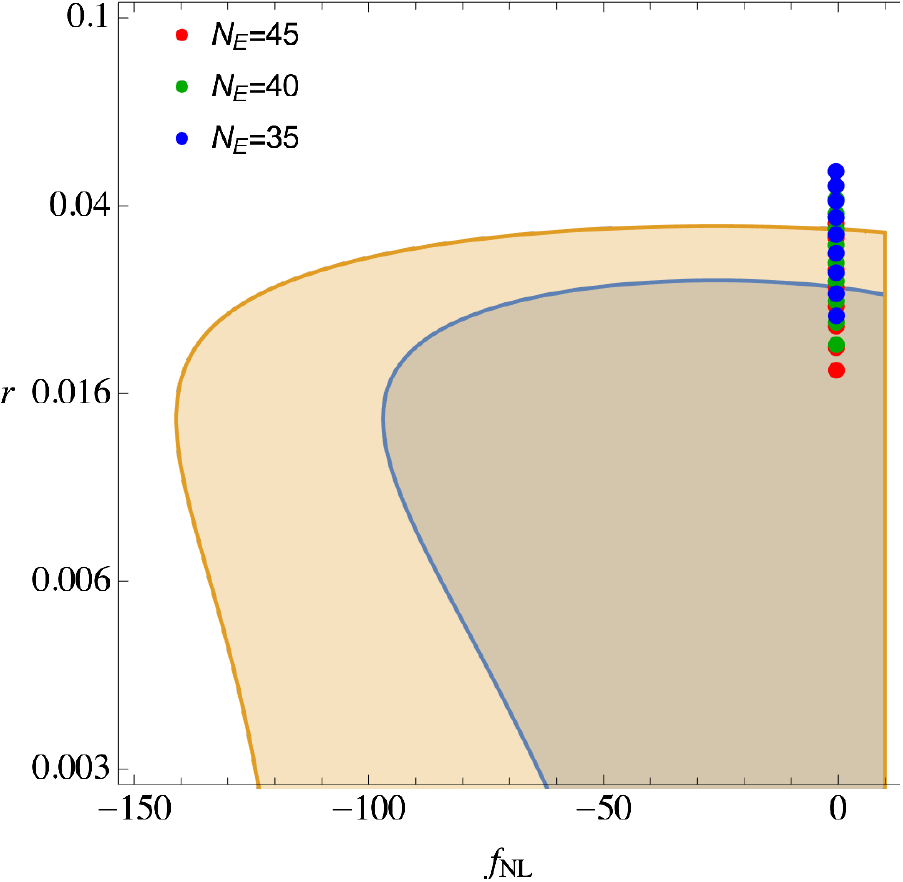}
    \includegraphics[scale=0.85]{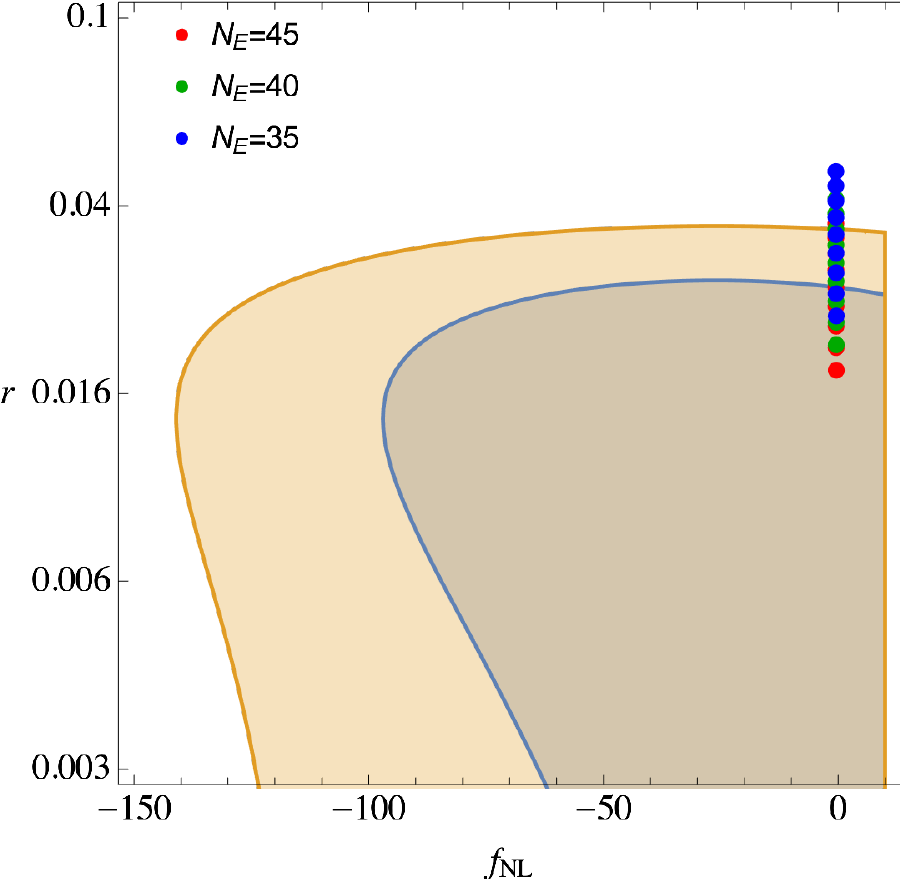}
	\caption{Tensor-to-scalar ratio $r$ v.s. equilateral non-Gaussianity $\fnl$ for double 
	monodromy with quartic (\emph{left}) and sextic (\emph{right}) kinetic terms.
    The parameters are the same as in Fig. \ref{fig:nsrplot}.
    Because of the lower bound on $c_s$ in either model ($c_s^2 > 1/3$ for quartic, 
    $c_s^2 > 1/5$ for sextic), the non-Gaussianity is very small.}
	\label{fig:rvsfnl}
\end{figure}

\begin{figure}[thb]
    \centering
    \includegraphics[scale=0.85]{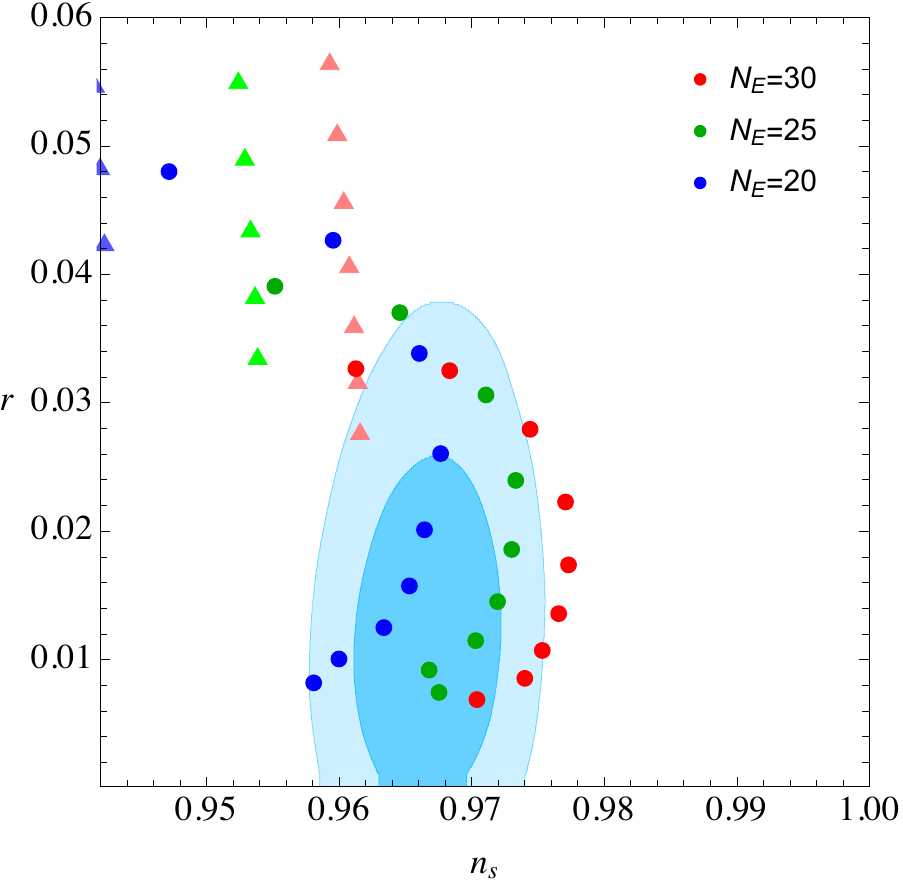}
    \includegraphics[scale=0.85]{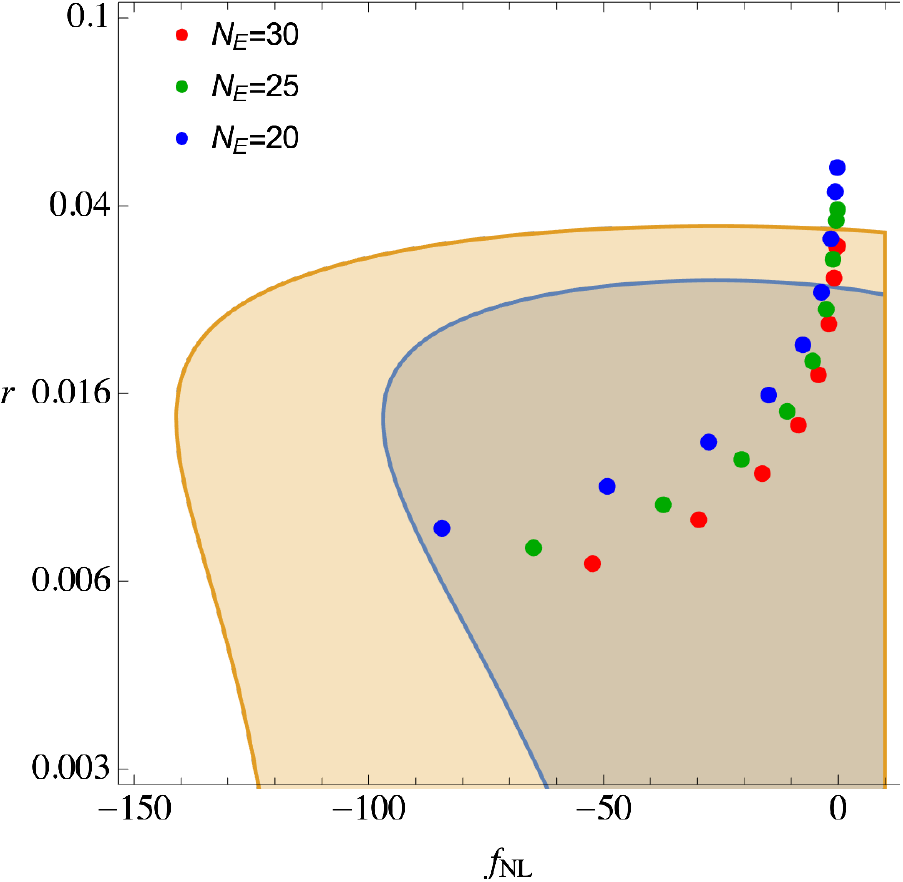}
    \caption{Tensor-to-scalar ratio $r$ versus spectral index $n_s$ (\emph{left}) and $r$ versus 
    equilateral non-Gaussianity $\fnl$ (\emph{right}) for double monodromy with DBI kinetic terms.
    We take $\mu_1 = 0.01$ for the potential, vary $0.1 \leq p \leq 0.5$ and fix $M = 0.23$.
    In the left panel, the triangles denote solutions with the canonical kinetic term, 
    while the dots are solutions for the DBI kinetic term.
    We note that, since $c_s$ is not bounded from below, $r$ can be much lower in the 
    DBI model with respect to quartic and sextic kinetic terms.
    Equilateral non-Gaussianity shows the expected inverse correlation with $r$, 
    but it is within the current experimental bounds.}
\label{fig:dbi}
\end{figure}

As in weak coupling, the amplitude perturbation is controlled by horizon scale when $\lambda^{-1} \simeq H$. 
The Gaussian curvature perturbation is determined by folding this scale with the time translation breaking scale. 
In contrast to the weak coupling regime, where 
$\sqrt{\dot{\varphi}} = (2 \Mpl^2 \dot{H})^{1/4} = \eps_H^{1/4} \sqrt{2 \Mpl H} \gg H$, 
in the strong coupling regime with large derivative terms one finds that 
this scale is $f_{\pi}^4 = 2 \Mpl^2 \dot{H} c_s$. Note, 
that by Eq. (\ref{kinflation}) in this regime $\Mpl^2 H^2 \simeq \frac{M^4}{16\pi^2}$.
The Gaussian scalar power spectrum is $\Delta_{\calR}^2 \propto (H/f_{\pi})^2$. 
 
The leading non-Gaussianities in this regime come from the 
3-point function. There are two operators in (\ref{Lpi}) which source them. However
since by naturalness and na\"ive dimensional analysis arguments all the derivatives are comparable, 
$c_3 \simeq c_s^2 < 1$, and so the non-Gaussianities generated by $\sim c_3$ term are subleading. 
Thus the amplitude of the cubic non-Gaussianities reduces to a single 
narrow strip \cite{DAmico:2017cda}. Using the perturbation potential 
\be
\langle \Phi_{\vk_1} \Phi_{\vk_2} \Phi_{\vk_3} \rangle
= (2 \pi)^3 \delta_D(\vk_1+\vk_2+\vk_3)
\frac{6 \Delta_{\Phi}^2}{(k_1+k_2+k_3)^3} \[ f_{\rm NL}^{(1)} F_1(k_1, k_2, k_3) 
+ f_{\rm NL}^{(2)} F_2(k_1, k_2, k_3) \] \, ,
\ee
we find that the $F_1$ and $F_2$ are induced by the $\dot{\pi} (\de \pi)^2$ and 
$\dot{\pi}^3$ operators, respectively~\cite{Chen:2006nt,Senatore:2009gt,Ade:2013ydc}.
Explicitly, we define
\begin{equation}
    K_1 = k_1 + k_2 + k_3 \, , \quad
    K_2 = (k_1 k_2 + k_2 k_3 + k_3 k_1)^{1/2} \, , \quad
    K_3 = (k_1 k_2 k_3)^{1/3} \, ,
\end{equation}
in terms of which
\ba
    &&F_1 = - \frac{9}{17} \frac{12 K_3^6 - 4 K_1 K_2^2 K_3^3 - 4 K_1^4 K_2^2 + 11 K_1^3 K_3^3 - 3 K_1^4  K_2^2 + K_1^6}{K_3^9} \, , \nonumber \\
    &&~~~~~~~~~~~~~~~~~~~~~~~~~~~~~~ F_2 = \frac{27}{k_1 k_2 k_3} \, .
\ea
The term $\propto F_1$ describes equilateral non-Gaussianities, since the momentum dependence is 
such that it is maximized when all three momenta are equal, while the term $\propto F_2$ shows an important contribution also on flattened triangles.
CMB analyses~\cite{Senatore:2009gt,Ade:2013ydc} are commonly done in terms of two templates, the equilateral one which is very similar to $F_1$, and an orthogonal one which is a linear combination of $F_1$ and $F_2$.
In terms of the Lagrangian parameters, the $f_{NL}$ coefficients are
\be
f_{NL}^{(1)} = - \frac{85}{324} \(\frac{1}{c_s^2}-1 \) \, , \qquad 
f_{NL}^{(2)} = - \frac{10}{243}
\(1-c_s^2 \) \(\frac{3}{2} + c_3 \) \, .
\label{eq:fnldef}
\ee
For our quartic and sextic models, $c_3$ in eq.~\eqref{eq:fnldef} is $-3$ and $-4$, respectively, so $f_{\rm NL}^{(2)}$ is very small with respect to $f_{\rm NL}^{(1)}$, by a factor $c_s^2$; for DBI, $f_{\rm NL}^{(1)} = - (17 / 4) f_{\rm NL}^{(2)}$, and they have the same dependence on $c_s^2$.
For definiteness, we show $f_{\rm NL}^{(1)}$ in Figs.~\ref{fig:rvsfnl} and \ref{fig:dbi}.
The experiments are already constraining the phase space of  even the strongly coupled theory.
From Planck~\cite{Planck:2019kim}, marginalizing over $c_3$, one gets the bound $c_s \ga 0.21$. This yields 
$X \la M^4/10$. Thus the theory should be in strong coupling to prevent too large $r$, but it cannot be arbitrarily 
strongly coupled to satisfy the bound on non-Gaussianities. Hence the tensor power cannot be arbitrarily weak. 

In Fig. \ref{fig:nsrplot} and \ref{fig:rvsfnl} we show the predictions for the observables 
$r$, $n_s$ and $\fnl$, for our potential Eq. (\ref{eq:Vdoublesin}) and kinetic terms with either quartic or 
sextic higher derivative terms, in a regime in which these higher derivative corrections 
are important. In Fig. \ref{fig:dbi}, we show the predictions for our double mondromy potential 
with a DBI kinetic term, $K(X) = M^4 \( 1 - \sqrt{1 - 2 X /M^4}\)$ which produces an 
infinite series of higher derivative corrections. In comparing to the results of \cite{Pedro:2019klo} it 
is important to note that there the higher derivatives arise from integrating out a heavy scalar generating 
an infinite series of higher derivative operators. In the particular example displayed in Fig. 5
in \cite{Pedro:2019klo}, this infinite series re-sums into a cosine potential in $V$ and a cosine 
dependence in $c_s$. This produces the strong reddening of $n_s$ with decreasing $c_s$ visible 
in the blue down-left curving band of Fig. 5 of \cite{Pedro:2019klo}. In our first example we 
only keep a quartic and/or sextic higher derivatives in the kinetic term, hence the reddening of 
$n_s$ remains much milder for our case. Moreover, with decreasing $c_s$ and thus increasing 
effects of the higher derivative  operators, we get pushed out of the quadratic part of the scalar 
potential onto the flatter-monomial `wings' which causes a blue-shift of $n_s$ partially offsetting 
the reddening effects of the higher derivatives themselves. Meanwhile, in the example of Fig. 5 
in \cite{Pedro:2019klo} decreasing $c_s$ and thus increasing effects of the higher derivatives 
pushes one towards the hill-top of a cosine potential which by itself causes additional reddening of $n_s$.

For the quartic and sextic kinetic terms, we observe that for powers $p \ga 0.1$, we get small non-Gaussianities, 
tensor to scalar ratio in the range $r \ga 0.015$, and $0.96 \la n_s \la 0.97$
for the first stage of inflation which ends after $35 - 45$ efolds.
For the DBI kinetic term, we get $r \ga 0.006$ and larger $\fnl$ for the 
first stage of inflation ending after $20-30$ efolds.
Summarizing, with non-standard kinetic terms the model remains a 
fully viable fit of the sky at the CMB scales, but with predictions in the reach of the 
near future cosmological observations, such as the next installment of BICEP, or LiteBIRD. 
This makes our natural monodromy models an excellent benchmark for future observations.
Further, the mechanism of nonperturbative generation 
of chiral tensors, using vector tachyon instability \cite{pinky,evazald,DAmico:2021vka}, 
which we review below in this new context, 
remains operational and can yield additional gravity wave signals at shorter scales.

\section{F\"ur LISA, an Encore, with Other Instruments Too}

We noted in \cite{DAmico:2021vka} that axion-like inflatons often couple to $U(1)$ gauge fields via the standard 
dimension-5 operators $\propto \phi_1  F_{\mu \nu} \tilde{F}^{\mu \nu}/2$.
A simple example is to start with a $4$-form field strength in 
$11D$ Sugra, with Chern-Simons self-couplings, and dimensionally reduce it on some 
toroidal compactification. Ignoring the moduli fields the resulting $4D$ effective action will include
light axions coupled to dark $U(1)$s since  
\ba
-F_{abcd}^2+ \epsilon_{a_1 \ldots a_{11}} A^{a_1 \ldots} 
F^{a_4 \ldots} F^{a_8 \ldots a_{11}} &\ni& - F^2_{\mu\nu\lambda\sigma} 
- (\partial \phi_1)^2 - \mu \phi_1 \epsilon_{\mu\nu\lambda\sigma} F^{\mu\nu\lambda\sigma} \nonumber \\
&& - \sum_k F^2_{\mu\nu \, (k)} - 
\frac{\phi_1}{f_\phi} \sum_{k,l} \epsilon_{\mu\nu\lambda\sigma} F^{\mu\nu}{}_{(k)}  F^{\lambda\sigma}{}_{(l)} \, .
\label{4formterms}
\ea
Here the first line lists the $4$-form-axion sector, which thanks to the monodromy coupling is massive but light,
and the second describes the mixing of the axion with dark $U(1)$s. If for simplicity we take 
only one coupling to be nonzero, we can model it with the canonically normalized $4D$ dimension-5 operator
\begin{equation}
    \calL_{\rm int} = - \sqrt{-g} \frac{\phi_1}{4 f_\phi} F_{\mu \nu} \tilde{F}^{\mu \nu} \, ,
\label{eq:coupling}
\end{equation}
where $f_\phi$ is sub-Planckian. This scale is generically $\simeq M_{GUT}$ 
(see e.g.~\cite{Banks:2003sx,Svrcek:2006yi}). A rolling axion triggers the 
tachyonic instability of one circular polarization of the gauge field \cite{nkaloper}, 
whose exponential production both backreacts on the inflaton and 
produces scalar and tensor perturbations \cite{pinky,evazald}. Details of the backreaction
were examined in \cite{Barnaby:2011qe,Linde:2012bt}, and a 
very comprehensive analysis of these effects was 
provided recently in \cite{Domcke:2020zez}, with the results of the non-perturbative 
treatment recently cross-verified in~\cite{Gorbar:2021rlt} 
using a completely different gradient expansion formalism down to numerically identical 
predictions of the GW signal including resonance-induced 
peaked fine structure. The details of the dynamics are given in \cite{Domcke:2020zez,DAmico:2021vka}
and we will not repeat them here. We should mention the other works 
which have recently explored bumpy early universe
dynamics to generate relic gravity waves 
\cite{Ragavendra:2020sop,Fumagalli:2020nvq,Anguelova:2020nzl,Cai:2020qpu,
Dalianis:2021iig,Fumagalli:2021mpc,Cui:2021are}. 

\begin{figure}[ht]
    \centering
    \includegraphics[scale=0.78]{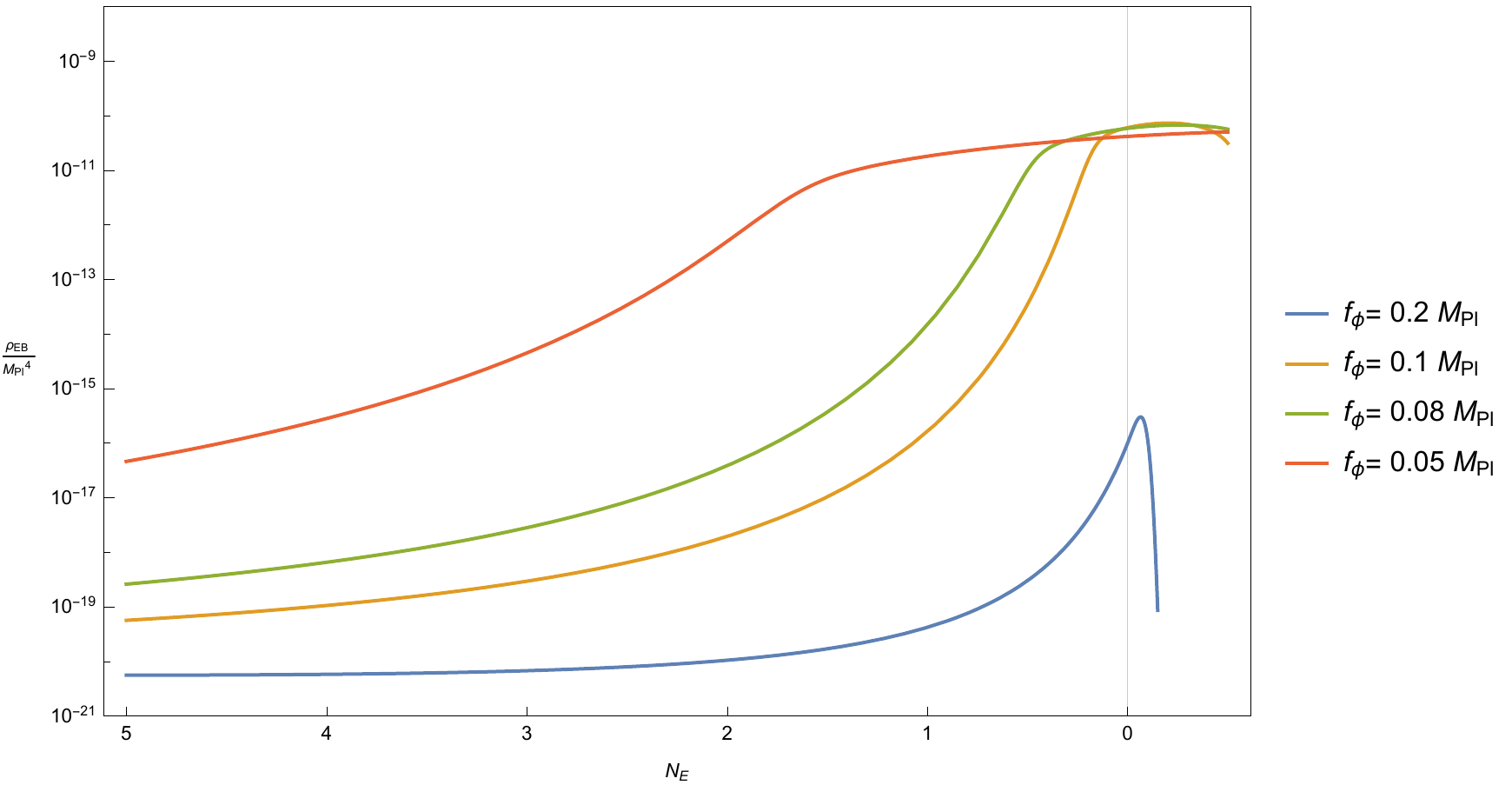}
    \caption{Evolution at the end of the first inflationary stage of the 
    energy density in gauge fields. 
    We denote by $N_e$ the number of efolds before the end of the first 
    stage of inflation, and we normalize the energy density to $\Mpl^4$.
    The parameters of the potential are fixed for convenience to $M_1^4 = 2 \times 10^{-9} \Mpl^4$, $\mu_1 = \Mpl$, 
    $p = 0.2$; at the CMB scales they lead to similar results as $\mu_1 = 0.1 \Mpl$, which fit CMB perfectly. 
    We show results for different couplings $f_{\phi}$ as shown in the legend.
    The vertical line denotes the end of inflation, and the solutions we show are not completely reliable there. 
    Note that the contributions of the vector field are very small until the very end of inflation, 
    when $\phi_1$ moves the fastest. As a result the U(1) production does not affect CMB significantly.}
\label{fig:rhoEB}
\end{figure}
We imagine this axion to be the inflaton dominating the first stage
of inflation in the double monodromy inflation of the previous section, 
which in addition to the simple quadratic potential
(given in the dual $4$-form frame by the first line of Eq. (\ref{4formterms})), 
also includes additional corrections, that combine
into the potential (\ref{eq:Vdouble}) and the higher derivative operators (\ref{correctionsscalarnorm}). 
Thus early on during the 
first stage of inflation, the dynamics described in the previous section
yields a suppressed tensor-to-scalar ratio and weak non-Gaussianities, while matching the scalar CMB spectrum.
By the end of this stage, however, the inflaton drops out of 
the strong coupling, and inflation continues for a few more efolds in weak coupling. This last epoch of the first
stage of inflation is described by the terms of Eq. (\ref{4formterms}). Hence 
the analysis of sub-CMB gravity wave generation of 
\cite{DAmico:2021vka}, describing how the tachyonic instability in the 
dark $U(1)$ is triggered by (\ref{eq:coupling}) goes through
unabated. Hence, a $U(1)$ chirality is cranked up and it in turn sources 
chiral gravity waves.  The amplification is quite efficient, 
although it is bounded by inflaton kinetic energy and scale of the first stage of 
inflation near its end. Assuming that the potential 
(\ref{eq:Vdoublesin}) remains valid all the way to the end of the weakly coupled epoch of the first stage of inflation,
the energy density transferred to the tachyonic chirality of the dark $U(1)$ can 
be calculated numerically, as presented in 
Fig. \ref{fig:rhoEB}.

We should stress that some of these modes will reenter the horizon during the intermediate matter-dominated stage, 
interrupting the two stages of inflation. Hence they would dilute by expansion during that epoch.
However as long as the interruption is short the dilution is weak. We can estimate it as follows: 
for subhorizon wavelength modes, the amplitude 
goes as $1/\lambda \sim 1/a$, where $a$ is the scale factor. 
Since these modes are harmonic oscillators, their power, 
by virial theorem, is given by the square of frequency $\times$ the amplitude. 
Hence the suppression factor will be at most $(a_1/a_2)^4$ where $a_1$ is the
scale factor at the end of stage 1 of inflation, and $a_2$ the scale factor at the mode re-freezing 
after the start of the next stage inflation. This factor is largest for the shortest wavelength mode 
at the end of stage 1, for which $k/a_1 \sim H_1$. Because the mode freeze-out yields $k/a_2 \simeq H_2$, 
this yields $a_1/a_2 \la H_2/H_1$, and so for a short interruption the power suppression would be no more than 
a factor of 100 to 1000, which is why we ignored it here. A more precise calculation is warranted here,
both for the computation of the proper signal benchmarks and, crucially, since this also 
evades violations of the BBN bound on gravity 
waves, $\Omega_{GW} h^2 \lesssim 10^{-6}$ \cite{Caprini:2018mtu}.
For this reason, we sketch these uncertainties in the prediction, plotting 
a gray band, spanning two orders of magnitude, 
around the idealized prediction of the power 
emitted in GWs in Fig. \ref{fig:OmegaGW}, \ref{fig:OmegaGWvsN} below.

\begin{figure}[th]
    \centering
    \includegraphics[scale=0.78]{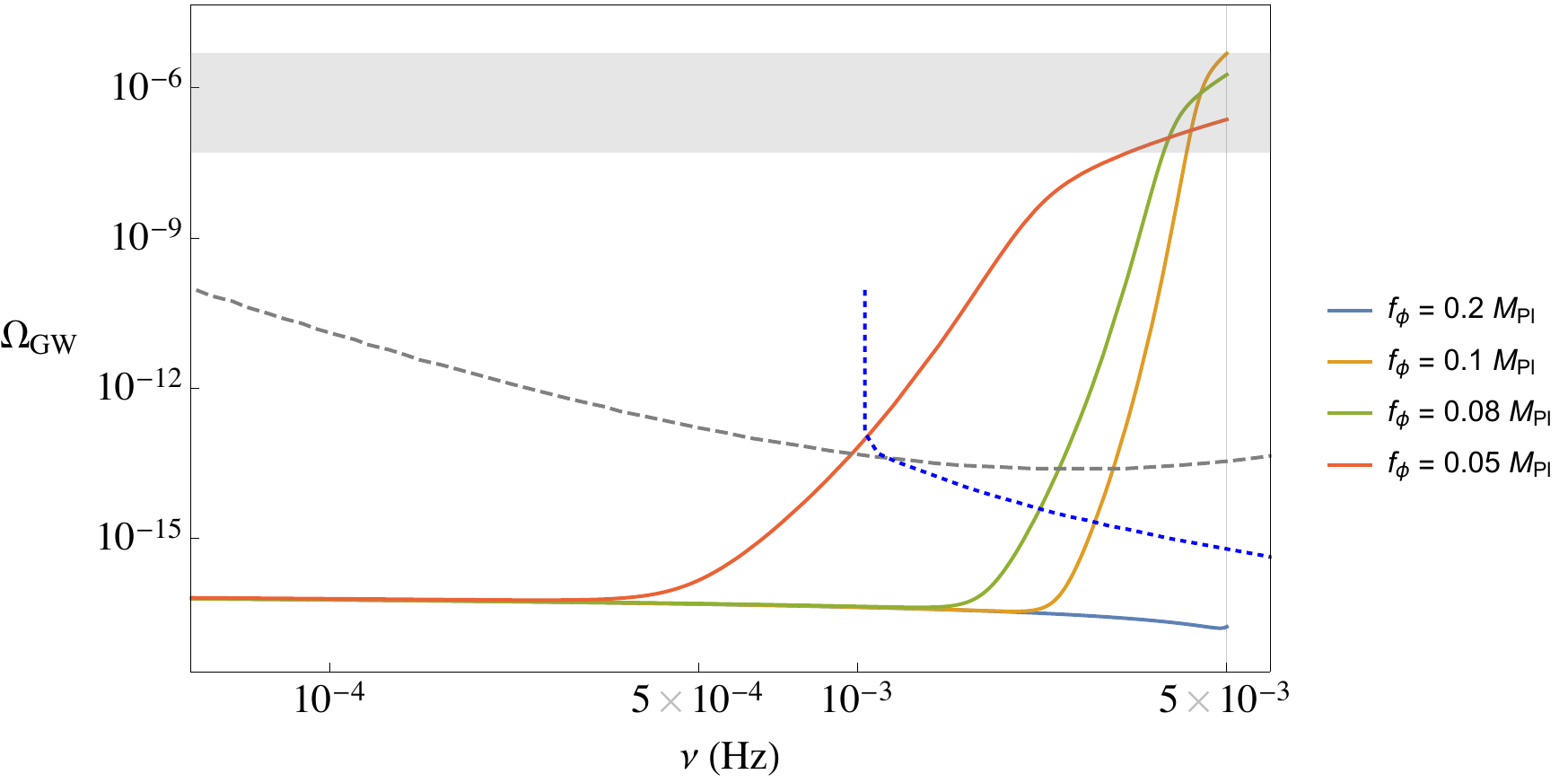}
    \caption{Abundance of gravitational waves as a function of frequency, setting $N_{\rm CMB}=35$.
    The dashed grey line is the sensitivity of LISA~\cite{Bartolo:2016ami} and the dotted blue line is the sensitivity of Big Bang Observer (BBO)~\cite{Yagi:2011wg}.
    Again, we use $M_1^4 = 2 \times 10^{-9} \Mpl$, $\mu_1 = \Mpl$, $p = 0.2$ for convenience, 
    and as before scan $f_{\phi}$ as shown in the legend. The vertical line again designates the end of inflation,
    beyond which a different approximation is needed. We plot maximal power here ignoring the suppression
    during the interruption. The gray band outlines the uncertainties due to this suppression, which
    when accounted for evade violations of the BBN bound on gravity waves.}
\label{fig:OmegaGW}
\end{figure}

In turn the tachyonic dark $U(1)$ modes source the stochastic gravity waves which are chiral, with the 
total abundance which can be estimated by \cite{Domcke:2017fix}
\begin{equation}
    \Om_{GW} \equiv \frac{\Om_{r,0}}{24} \Delta_T^2
    \simeq \frac{\Om_{r,0}}{12} \(\frac{H}{\pi \Mpl}\)^2
    \( 1 + 4.3 \cdot 10^{-7} \frac{H^2}{\Mpl^2 \xi^6} e^{4 \pi \xi} \) \, ,
    \label{abundance}
\end{equation}
where $\Om_{r,0} = 8.6 \cdot 10^{-5}$ is the radiation abundance today, $\xi = \frac{\dot{\phi}}{2 H f_\phi}$ and the terms are evaluated at horizon crossing.
The equation is valid for $\xi \gtrsim 3$. This is the sum of the two polarization of gravitational waves: the $"1"$ term in the parenthesis arises due to the usual metric fluctuations in de Sitter space, and includes the contributions
from both graviton helicities. The second term in parenthesis, involving the exponential amplification, receives contribution only from the helicity sourced by the `tachyonic' gauge field source, 
as evidenced by its dependence on $\xi$. By mode orthogonality in the linearized limit, the other graviton
helicity is not enhanced.

It is clear that a favorable realization of the scales can enhance the primordial GW spectrum dramatically.
To fix the physical wavelength of these modes, which allows us to see 
by which instrument they might be searched for, and
what their amplitude at the relevant wavelength is, 
we can rewrite $\Omega_{GW}$ as a function of 
the frequency observed at the present time. Since 
the comoving frequency is $\nu = k/(2 \pi)$, the frequency of the 
modes in terms of the number of efolds before the end of inflation
is given by
\begin{equation}
    N = N_{CMB} + \ln \frac{k_{\rm CMB}}{0.002 \Mpc^{-1}} - 44.9 - \ln \frac{\nu}{10^2 \Hz} \, ,
\end{equation}
where $k_{CMB} = 0.002 \Mpc^{-1}$ is the CMB pivot scale, and $N_{\rm CMB}$ is 
the number of efolds before the end of the first stage of inflation where the CMB scales froze out.
With this we can regraph the results of Fig. \ref{fig:rhoEB} in terms of the new independent variable $\nu$ and the
GW amplitude $\Omega_{GW}$. The results are presented in Fig. \ref{fig:OmegaGW}. 
Again the approximations which 
we employ are not completely reliable beyond the end of inflation, and noted above and in \cite{DAmico:2021vka}. 
However they remain a good indicator of the signal's power.
\begin{figure}[th]
    \centering
    \includegraphics[scale=0.78]{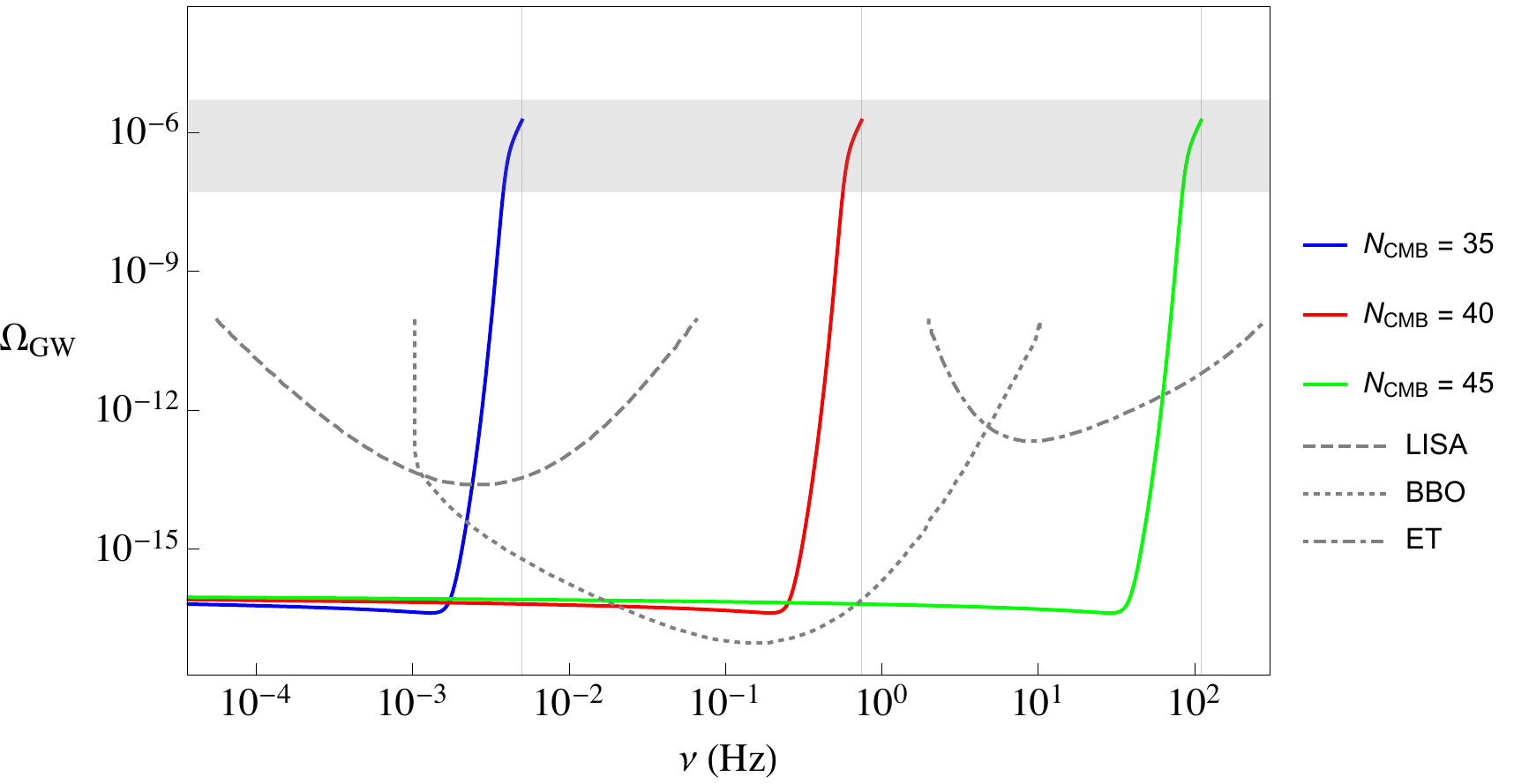}
    \caption{Abundance of gravitational waves as a function of frequency, setting 
    for different $N_{\rm CMB}=35, 40, 45$.
    We show the predicted bounds as a function of frequency for LISA~\cite{Bartolo:2016ami}, Big Bang 
    Observer (BBO)~\cite{Yagi:2011wg} and Einstein Telescope (ET)~\cite{Hild:2010id}.
    We use $M_1^4 = 2 \times 10^{-9} \Mpl$, $\mu_1 = \Mpl$, $p = 0.2$, $f_{\phi} = 0.08$. If $n_s$ is bluer, for 
    monodromy models this means longer (first stage of) inflation. If at the end of this stage chiral tensors are 
    generated, their frequency will be higher. The vertical line designates the end of inflation 
    for each curve, beyond which a different approximation is needed. Again the gray band outlines the 
    uncertainties in the prediction of the power due to the suppression at the interruption between the
    stages of rollercoaster inflation.}
\label{fig:OmegaGWvsN}
\end{figure}

Of course, as we have seen from the previous discussion, and showed manifestly in e.g. Fig. \ref{fig:nsrplot}, 
we can automatically satisfy the CMB bounds on $n_s$ and $r$ for a range of $N_{CMB}$. 
This means, that there are degeneracies in the evolution 
allowing for a good fit the CMB for a range of models, with different values of the 
pivot point $N_{CMB}$ -- i.e. with the ``CMB epoch"
of multistage inflation of varying duration. Therefore we find it interesting to fix $f_{\phi}$ and plot 
$\Omega_{GW}(\nu)$ for different values of $N_{CMB}$, which generates a
horizontal shift of the curves of Fig. \ref{fig:OmegaGW}.
This plot is in Fig. \ref{fig:OmegaGWvsN}, from which it is apparent that several experiments 
in the near future can explore the parameter space of our model. What's more is that such a 
mechanism might occur at the end of every accelerated stage of rollercoaster, due to the
inflatons being axion-like and possibly coupling to many different dark $U(1)$s, 
thus producing what we might dub a ``characteristic spectrum"
of rollercoaster models. If so then each spike, a different range of scales, 
might be simultaneously probed by one of the planned instruments.

The bottomline is that these modes may very well be out there for the 
GW instruments to discover. Depending on the specifics of the earliest stage of inflation
and its duration, their wavelength might be in the sweet spot of LISA~\cite{Bartolo:2016ami} or DECIGO/BBO~\cite{Yagi:2011wg}, or if the 
wavelength is longer (and the earlier stage of inflation shorter), SKA or NANOgrav 
(for a recent study of the measurement of the spectrum of primordial gravitational waves 
see~\cite{Campeti:2020xwn}). In either case, combining this with 
the bounds on $r$ at the CMB scales makes our 
double-coaster, and more generally multistage rollercoaster, extremely predictive and easy to confirm.

\section{Outro}

We have updated here our earlier model of double monodromy inflation with the inclusion 
of the general strong coupling-induced
irrelevant operators, which lead to additional suppression of the tensor to scalar ratio in inflation. 
As a result, the predictions of the model, which involves an early stage of inflation which is interrupted
by the first inflaton decay $25-40$ efolds after the beginning, are fully consistent with the most recent
BICEP/{\it Keck} bounds. The observables are $0.006 \la r \la 0.035$ with $0.96 \la n_s \la 0.97$ with
$f_{NL} \la {\cal O}(10)$. This makes the models very predictive since they can 
be constrained -- and possibly confirmed
and ruled out -- by the very near future observations. In addition the first inflaton 
could couple to a hidden sector $U(1)$ and lead to an enhanced production of vectors 
near the end of the first stage of inflation. These vectors 
in turn would source tensors at shorter wavelengths, leading to
additional signatures of the model that would correlate with the signatures in the CMB.

In closing we should finally mention that at this point however it may still be premature to take the $n_s-r$ bounds 
as very strong obstructions to inflationary models. Namely, although the bounds on $r$ are strong, the correlation
of $n_s$ and $r$ might be weaker than it seems. We believe the situation warrants a warning of sorts since the
$n_s - r$ correlation is commonly used to assess the likelihood of popular inflationary 
models (nice reviews of the observational constraints on theoretical models can be found in,
e.g. \cite{eva,Kallosh:2021mnu}). For example, one commonly encounters the ``potato plots" such as Fig. (5) 
in reference \cite{bicep}, or Fig. 4 in the subsequent paper \cite{tristram}, as well as our 
own Fig. \ref{fig:nsrplot} - \ref{fig:dbi}, which one may take to suggest there is very small remaining
parameter space for inflationary models to `squeeze in'. Such reasoning might be too quick, since 
there is another interesting possibility that can open up the space in the $n_s-r$ 
plane, which is quite curious although it might
seem somewhat extreme. 

The issue is that in all the figures plotting $r$ versus $n_s$ which we have shown so far, we have 
relied on the constraints on $n_s$ based on the CMB data analysis \emph{assuming} the $\Lambda$CDM model of
the late universe. However, for some years now, the measurements of the Hubble 
parameter using standard candles \cite{Bernal:2016gxb,Verde:2019ivm,Freedman:2019jwv,Riess:2021jrx} 
have shown discrepancy with the value of $H_0$ determined by 
Planck experiment, which is based on fitting the CMB data to 
the late $\Lambda$CDM cosmology \cite{Ade:2018gkx}.
To resolve the Hubble tension, many models have been proposed to date 
(for reviews see~\cite{DiValentino:2021izs,Schoneberg:2021qvd}). Among the most successful models
are the so called ``Early Dark Energy" proposals, which involve 
additional degrees of freedom before recombination 
\cite{Kamionkowski:1992mf,Doran:2000jt,Niedermann:2019olb,Niedermann:2020dwg}. 
Analyzing the CMB data in terms of these new models, a common feature is that the spectral index 
is bluer with respect to the $\Lambda$CDM value 
(see~\cite{DiValentino:2018zjj,Ye:2020btb,Ye:2021nej,Tanin:2020qjw,Takahashi:2021bti}). 
If it indeed turns out that the 
Hubble tension is real, and the most recent examinations indicate support for this option,
assessing the discrepancy to be at $5 \sigma$ at this time \cite{Riess:2021jrx}, 
the shift of $n_s$ will have very important implications for inflationary model constraints.

\begin{figure}[thb]
    \centering
    \includegraphics[scale=0.85]{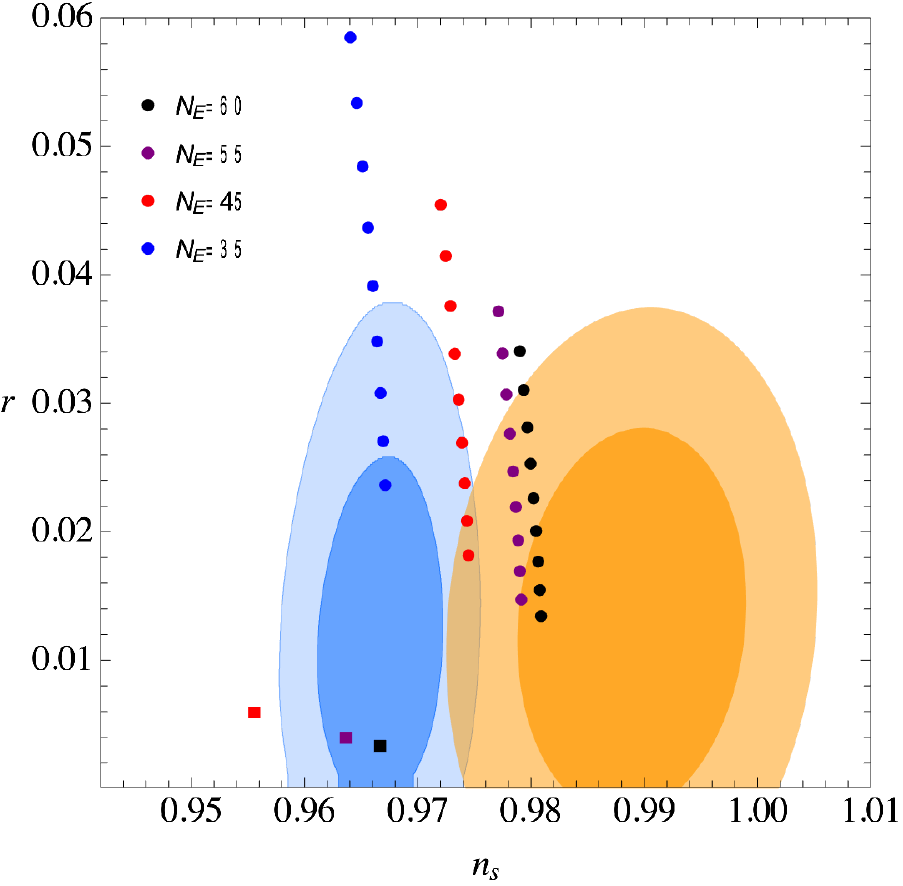}
    \caption{Tensor-to-scalar ratio $r$ versus spectral index $n_s$, for the $\Lambda$CDM model (blue) 
    and for the NEDE model (gold). To get the NEDE model constraint, we approximate the $\Lambda$CDM 
    contour as a bivariate Gaussian, and substitute the mean and error on $n_s$ by the ones got in the NEDE model. 
    This approximate procedure reproduces well a full analysis~\cite{Ye:2021nej}. The round dots are predictions of
    potentials (\ref{eq:Vdouble}) with $0.1 \leq p_1 \leq 0.5$, as before. The square dots are the predictions of the
    Starobinsky $R^2$ inflation \cite{Starobinsky:1980te}.}
\label{fig:ede_ns}
\end{figure}

We illustrate this in Fig. \ref{fig:ede_ns}, where we plot the contours obtained from the $\Lambda$CDM model, 
and the one we infer from the constraints on $n_s$ in the so-called NEDE model, 
presented in~\cite{Niedermann:2019olb,Niedermann:2020dwg}. 
This plot clearly shows how -- all of a sudden -- 
flattened monodromy models, with a long stage of inflation, and the 
power law behavior of, for example, $\sim \phi^{1/2}$ at large $\phi$, 
which were supposedly excluded by having too large an $n_s$, may end up being better 
candidates than, for example, the Starobinsky $R^2$. Of course,
it is too soon to claim this. But until the $H_0$ tension is resolved, it may also be too soon to claim the opposite. 
The point is that we need to resolve the issues which arose in the late universe cosmology before
we can get into the precision data confirming or ruling out inflationary models. 
With the ever better quality of data and an 
array of planned searches and tests in the very near future, this seems to be within reach. 

We therefore remain 
quite curious about what the future observations may yield.

\vskip.3cm

{\bf Acknowledgments}: 
We would like to thank V. Domcke, P. Graham, R. Kallosh, A. Lawrence, A. Linde, F. Niedermann, 
E. Silverstein, M. Sloth and T. Weigand
for many very useful discussions. NK is supported in part by the DOE Grant 
DE-SC0009999. AW is supported by the ERC Consolidator Grant STRINGFLATION 
under the HORIZON 2020 grant agreement no. 647995.

\bibliographystyle{utphys}
\bibliography{references}

\end{document}